%% file: bubbmerge_subm.tex
\documentclass[debug,overfull,figures]{epl}

\usepackage{rotating}
\usepackage{amsmath}
\usepackage{amssymb}
\usepackage{mathrsfs}
\usepackage{graphics}
\usepackage{graphicx}
\usepackage{epsf}
\usepackage{color}

\title{Bubble coalescence in breathing DNA: Two vicious \\ walkers in
opposite potentials}
\shorttitle{Bubble coalescence}

\author{Tom{\'a}{\v s} Novotn{\'y}\inst{1}\inst{2}, Jonas Nyvold
Pedersen\inst{3}, Tobias Ambj{\"o}rnsson\inst{4}\inst{5},\\ Mikael
Sonne Hansen\inst{6}, \and Ralf Metzler\inst{4}\inst{7}}
\shortauthor{T.~Novotn{\'y} \etal} \institute{\inst{1}
Nano-Science Center, University of Copenhagen
                    - Universitets\-par\-ken 5, 2100 Copenhagen, Denmark\\
           \inst{2} Dept.~of Condensed Matter Physics, Faculty of Mathematics
                    and Physics, Charles University - Ke Karlovu 5, 121 16
                    Prague, Czech Republic\\
           \inst{3} Mathematical Physics, Lund University - Box 118,
                    22100 Lund, Sweden\\
           \inst{4} NORDITA - Blegdamsvej 17, 2100 Copenhagen, Denmark\\
           \inst{5} Dept.~of Chemistry, Massachusetts Institute of Technology -
                    77 Massachusetts Ave, Cambridge, MA 02139, USA\\
           \inst{6} Dept.~of Mathematics, Technical University of
                    Denmark - Bldg. 303S, Matematiktorvet, 2800
                    Kgs. Lyngby, Denmark\\
           \inst{7} Physics Dept., University of Ottawa - 150 Louis Pasteur,
                    Ottawa, ON, K1N 6N5, Canada}


\begin{document}

\maketitle

\begin{abstract}
We investigate the coalescence of two DNA-bubbles initially located at weak
segments and separated by a more stable barrier region in a
designed construct of double-stranded DNA. The characteristic time
for bubble coalescence and the corresponding distribution are
derived, as well as the distribution of coalescence positions
along the barrier. Below the melting temperature, we find a
Kramers-type barrier crossing behaviour, while at high
temperatures, the bubble corners perform drift-diffusion towards
coalescence. The results are obtained by mapping the bubble
dynamics on the problem of two vicious walkers in opposite
potentials.
\end{abstract}

\section{Introduction}

The Watson-Crick double helix is the thermodynamically stable
state of double-stranded DNA in a wide range of temperatures and
salt conditions \cite{kornberg}. This stability is effected by
hydrogen bonds between the bases in individual base-pairs (bps),
and the stronger stacking interactions between nearest neighbour
pairs of bps. Driven by thermal fluctuations double-stranded DNA
can break apart, to form denaturation bubbles of flexible
single-stranded DNA \cite{poland}. Although rare, bp-opening
events expose active groups of DNA bases, that are otherwise
buried within the double helix. They are crucial for the
interaction with proteins and chemicals, and therefore for the
biological function of DNA. Bubble kinetics has been probed in NMR
studies \cite{gueron}, and the growth and reannealing dynamics of
individual bubbles has been measured in real time in single DNA
fluorescence correlation setups \cite{altan}.

\begin{figure}
\begin{center}
\scalebox{0.6}{\input{bubbsketch_new_thin.pstex_t}} \caption{{\bf
Left:} Schematic of the bubble coalescence setup in a designed DNA
construct. It is clamped at both ends and consists of two outer
soft zones (thin red lines) of lengths $N_L,N_R$ bps with
melting temperature $T_s$ and a stronger $N$-bps-long barrier zone
(thick blue lines) with $T_b>T_s$. a) All bps closed ($T<T_s<T_b$).
b) Soft zones open by raising the temperature above $T_s$.
b$_1$--b$_3$) Successive opening of the barrier driven mainly by
fluctuations ($T<T_b$) or drift ($T>T_b$) until coalescence. The
discrete coordinates $X,Y=0,\dots,N$ are defined as the positions
of the {\em interfaces} between the closed and broken bps. {\bf
Right:} Plot of the linear potentials experienced by the
respective bubble interfaces in the case $T<T_b\ (f<0)$ in terms
of the dimensionless quantities $x,y,f$ (see text).}
\label{sketch}
\end{center}
\end{figure}
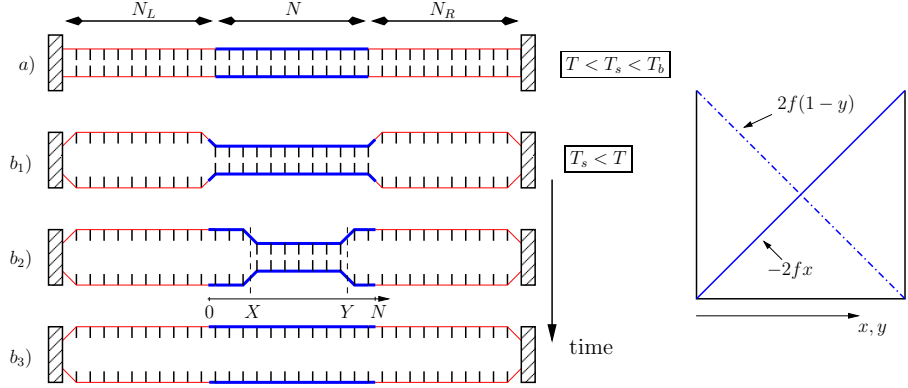

The delicate sensitivity of bubble dynamics and, therefore, bubble
nucleation to the local DNA bp-sequence \cite{prl,choi} suggests a
new method to use single molecule tools to obtain independently
DNA stability parameters, as sketched in figure \ref{sketch}. A
short stretch of DNA, clamped at both ends, is designed such that
two soft zones consisting of weaker AT-bps
are separated by a more stable
barrier region rich in GC bps. For simplicity, we assume that both
soft zones and barrier are homopolymers with a bp-dissociation
free energy $\varepsilon'$ and $\varepsilon$, respectively, and,
in accordance with the experimental findings of reference
\cite{altan}, we neglect secondary structure formation in the
barrier zone. At temperatures higher than the melting temperature
$T_s$ of the soft zones but still lower than the melting
temperature $T_b$ of the barrier region, thermal fluctuations will
gradually dissociate the barrier, until the two bubbles coalesce.
We also study the case when the system is prepared as above and
then $T$ suddenly increased such that $T>T_b>T_s$ so that the
system is driven towards coalescence. In both cases the two
boundaries between bubbles and barrier perform a (biased) random
walk in opposite free energy potentials. The quantities of
interest are the bubble coalescence time and position. It turns
out that this problem can be mapped on a previously unsolved case
of two vicious walkers in opposite potentials in one dimension
and, therefore, is of interest by itself.

\section{Formulation of the problem}

The statistical weight of the construct before coalescence
$\mathscr{Z}_{X,Y}=\big(\xi
e^{N_L\beta\varepsilon'}\big)e^{(X-Y+N)
\beta\varepsilon}\big(\xi e^{N_R\beta\varepsilon'}\big)$ at
$T_b>T>T_s$ involves a cooperativity (ring) factor $\xi\approx
10^{-5}$ for each bubble, and a Boltzmann factor for each broken
bp with free energies $\varepsilon'>0$ and $\varepsilon<0$,
compare reference \cite{prl}.\footnote{We denote a
stable state by a negative free energy.} Upon
coalescence, the boundary free energy corresponding to one factor
$\xi$ is released, $\mathscr{Z}_{\mathrm{coal}}=\xi e^{(N_L+N_R)
\beta\varepsilon'+N\beta\varepsilon}$, stabilizing the system
against immediate transition back to a two-bubble state. It should
therefore be possible to experimentally detect the coalescence. In
our theoretical approach, this corresponds to an absorbing
boundary for the random walking interfaces at $X=Y$.

\begin{figure}
\twoimages[height=5.0cm]{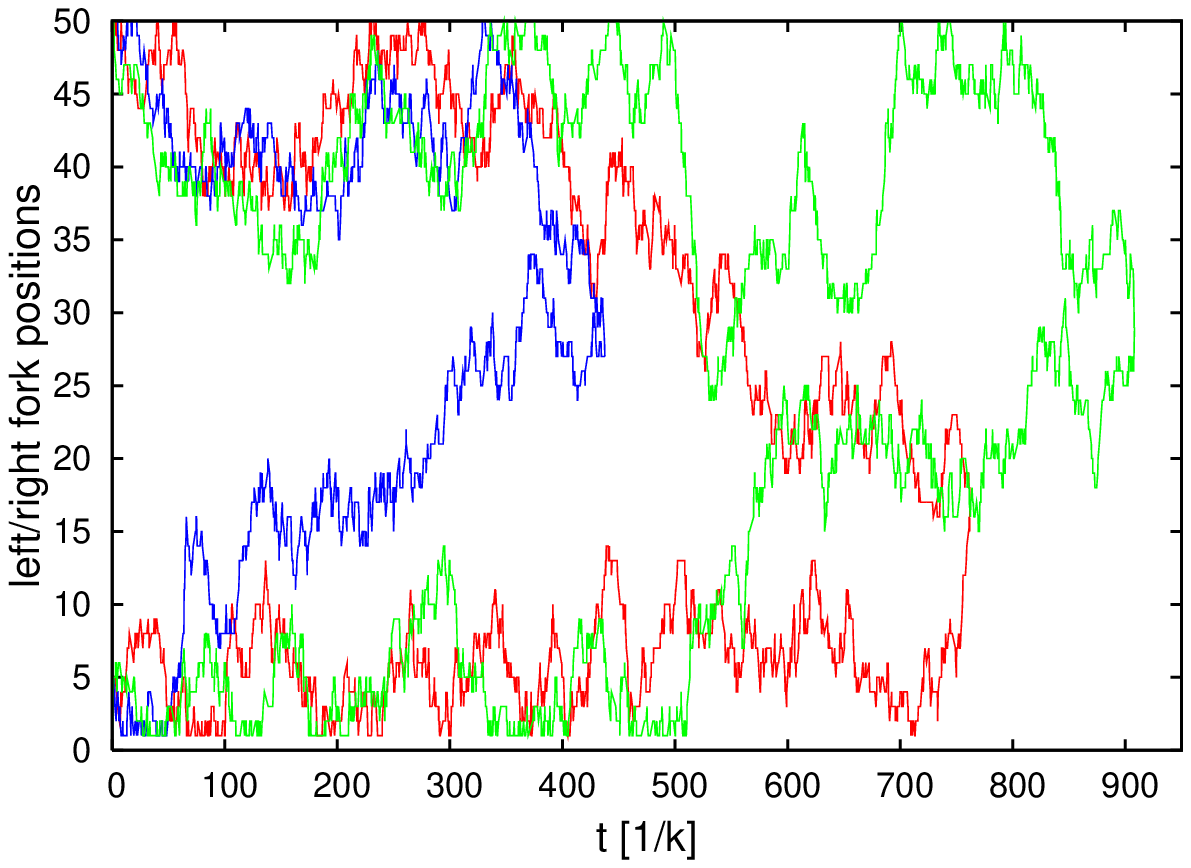}{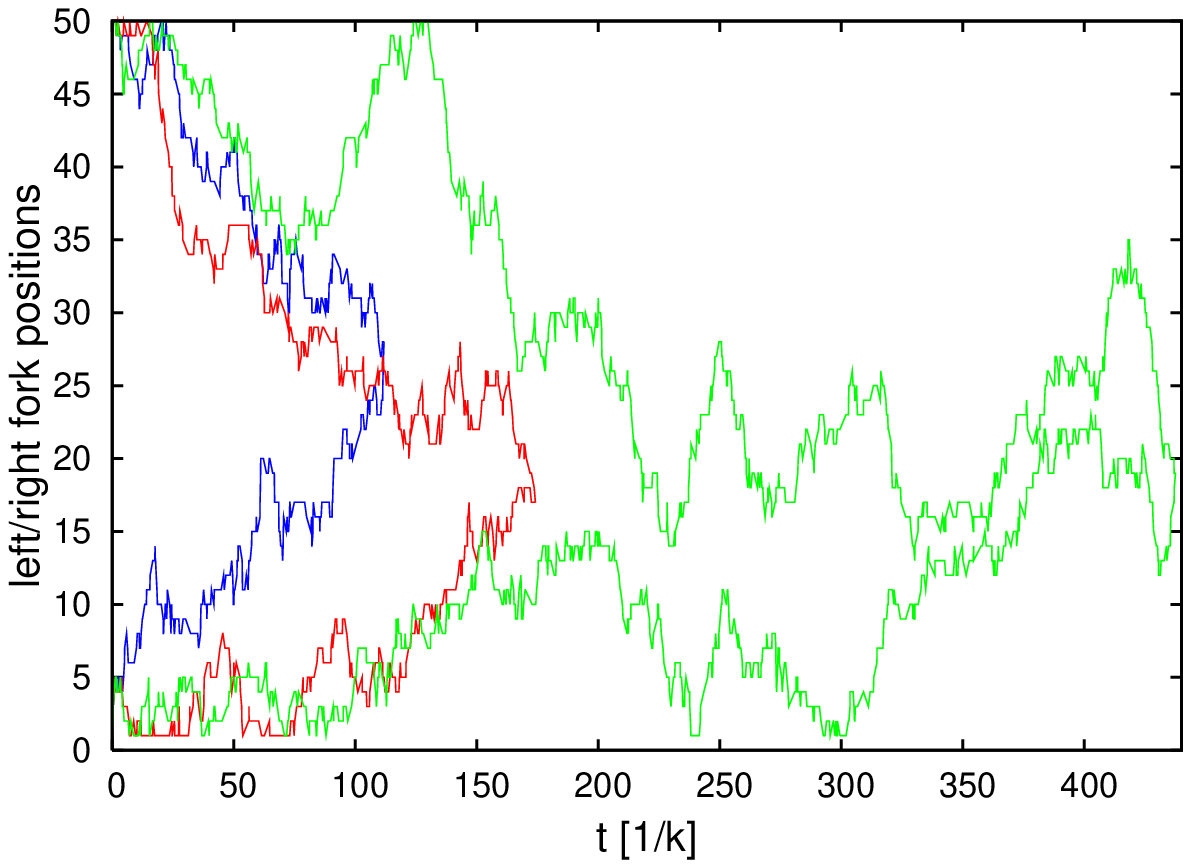}
\caption{Stochastic trajectories of the random walk of the two
fork positions encompassing the double-helical bridge of $N=50$
bps between the vicinal bubbles for $T_s <T<T_b$ with
$u=\exp(\beta\varepsilon)=0.98$ (left) and $T>T_b>T_s$, $u=1.10$
(right). Note that below the melting temperature $T_b$ of the
barrier zone the trajectories have the tendency to move toward the
reflecting boundaries at the corners of the soft zone and the
coalescence takes longer time due to the presence of the potential
barrier. On the other hand, above $T_b$ the trajectories show a
tendency to move fast toward the centre of the barrier zone, as
expected from the funnelling nature of the positive force directed
toward the middle.} \label{trajs}
\end{figure}

The dynamics of this, intrinsically discrete, system can be
modelled by methods developed previously, namely, the stochastic
Gillespie scheme \cite{suman}, which allows one to obtain single
bubble trajectories and thereby the access to the noise relevant
in single molecule experiments, and the master equation approach
\cite{jpc,pre,prl}. As detailed in reference \cite{jonas_long},
the explicit sequence of bps is included into these approaches.
Typical examples of individual trajectories resulting from the
Gillespie scheme are displayed in figure \ref{trajs}, where traces
of the two interfaces (``forks'') cornering the barrier region are shown.
Bubble coalescence terminates each pair of trajectories.

Both the Gillespie scheme and the master equation are based on a
specific choice for the rate constants, $\mathsf{t}^{(+)-}(Q)$, of
(un)zipping the bp at the position $Q=X,Y$ of one of the zipping
forks where double-stranded DNA branches into two strands of
single-stranded DNA. As discussed in detail in references
\cite{jpc,prl,pre,jonas_long}, we define
\begin{equation}
\label{rates} \mathsf{t}^-=k/2;
\,\,\mathsf{t}^+(Q)=ku(Q)s(m)/2,\,\,s(m)=\left\{(m+1)/(m+2)
\right\}^c ,
\end{equation}
together with the appropriate boundary conditions. Here,
$\mathsf{t}^-$ is the position-independent zipping rate $k/2$ for
a bp at the zipping fork, while the bp-unzipping at location $Q$
involves the Boltzmann factor $u(Q)= \exp(\beta\varepsilon(Q))$
for disrupting the bp at $Q$, as well as the factor $s(m)$ that
stems from the entropy loss of forming a closed polymer ring. The
$s$-factor depends on the size $m$ of the bubble cornered by the
respective bp, and $c\approx 1.76$ is the critical exponent. In
our illustrations and analytic approach we use the reflecting
boundary conditions at the edges of the barrier regions, i.e., we
assume $\beta\varepsilon'\gg1$ so that the soft zones are always
open.

The rates $\mathsf{t}^{\pm}$ define the transfer matrix
$\mathbb{W}$, that governs the random walk of the zipping forks at
either end of the barrier region. The coalescence dynamics is then
quantified by the probability distribution $P_D(X,Y,t)$ to find
the left and right zipping forks at positions $X$ and $Y$, as
controlled by the master equation $\partial P_D(X,Y,t)/\partial
t=\mathbb{W}P_D(X,Y,t)$. Solution of this equation by either
method, as detailed elsewhere \cite{prl,jpc,suman,jonas_long}, yields
the quantities of interest such as the mean coalescence time or
distribution of the coalescence position.

\section{Continuous description and semi-analytic solution}

Both the master equation and Gillespie approaches can be used for
arbitrary bp sequences and their usage is straightforward,
however, they both have their limitations. The Gillespie algorithm
performs badly in the case of a relatively strong barrier
($\beta\varepsilon\ll-1$) when the coalescence time grows
exponentially. This demands exponentially increasing simulation
time. The master equation is becoming numerically prohibitive with
increasing size of the barrier. On a regular PC the length of
about $N\approx 100$ bps is close to the numerical limit of our
master equation approach.

For the designed construct we can find an efficient
semi-analytical solution for its dynamics. Due to the large length
of the soft zones $N_{L,R}\gg 1$ we can well approximate the
bubble entropic factors $s(m)=\left\{(m+1)/(m+2) \right\}^c$ by
$1$. Furthermore, for the barrier length $N \gtrsim 50$ we can
resort to a continuous description of the fork positions.
%\footnote{The
%continuous description is actually not necessary for the usage of
%the following trick and is only employed for convenience. We could
%in principle analogously repeat the whole procedure within the
%discrete description in the spirit of reference \cite{bicout}. The
%linear potential is, however, still crucial for the validity of
%the method.}
The discrete master equation can then be rewritten as a bivariate
Fokker-Planck equation \cite{vankampen,risken} for the random walk
of the two bubble-barrier boundaries characterized by the
probability density function (PDF) $P(x,y,t)$ of finding the
boundaries at coordinates $x=X/N$ and $y=Y/N$ at a given rescaled
time $t$
\begin{equation}\label{fpe}
\frac{\partial}{\partial
t}P(x,y,t)=\left(\left[\frac{\partial^2}{\partial
x^2}+\frac{\partial^2}{\partial
y^2}\right]-2f\frac{\partial}{\partial x}+
2f\frac{\partial}{\partial y}\right)P(x,y,t),
\end{equation}
with the dimensionless force $f=N(u-1)/(1+u)$ and time rescaled by
$k(1+u)/2N^2$. Equation (\ref{fpe}) is completed by the initial
condition $P(x,y,0)=\delta(x-x_0)\delta(y-y_0)$ and the reflecting
boundary conditions (the bubbles in the soft zones are assumed to
be open at all times)$\left.(\partial/\partial x-2
f)P(x,y,t)\right|_{x=0}= \left.(\partial/\partial y+2
f)P(x,y,t)\right|_{y=1}=0$. Moreover, we impose the absorbing
boundary condition $P(x,x,t)=0$. This defines the vicious walker
property \cite{fisher}, terminating the process when the two
walkers meet.

The dynamics of any number of vicious walkers in an arbitrary but
{\em common} potential was studied previously \cite{bray},
obtaining the solution by antisymmetrization of the single-walker
evolutions (Green's functions). This method, however, already
fails for two walkers in two \emph{different\/} potentials. Of
course, equation (\ref{fpe}) can be solved numerically as a
(2+1)-dimensional partial differential equation, but it is clear
that expressing the solution in terms of single-walker Green's
functions would be very advantageous. It reduces the dimension of
the problem by 1 and even enables us to make exact analytical
statements about some aspects of the solution.

We show that for the particular problem at hand we can find the
solution to equation (\ref{fpe}) by antisymmetrization of a
suitable auxiliary single-walker Green's function. This is
achieved by the similarity transform of the Fokker-Planck operator
into a Hermitian operator (see reference \cite{risken}, chapter 5)
by using $\exp(\mp f(x-x_0))\left(\tfrac{\partial^2}{\partial
x^2}\mp2f \tfrac{\partial}{\partial x}\right)\exp(\pm
f(x-x_0))=\tfrac{\partial^2}{\partial x^2}-f^2$ which turns both
Fokker-Planck operators into the \emph{same\/} Hermitian one.
Together with the freedom to choose the same boundary conditions
for both walkers\footnote{Each walker can reach only one boundary
since the other one is blocked by the second walker. Therefore, we
can choose the remaining boundary condition at will and, in
particular, to be that of the second walker.} we end up with
identical differential equations for both walkers and, thus, the
antisymmetrization procedure can be used for the Hermitian version
of the problem. This finally results in
\begin{equation}\label{sol}
P(x,y,t)=e^{f(x-y-x_0+y_0)}\Big(p(x,t|x_0)p(y,t|y_0)-p(y,t|x_0)p(x,t|y_0)
\Big),
\end{equation}
where the auxiliary single-walker Green's function $p(x,t|x_0)$
satisfies the following 1D equation
\begin{equation}
\label{fde1} \frac{\partial p(x,t|x_0)}{\partial
t}=\left(\frac{\partial^2}{\partial x^2}-f^2 \right)p(x,t|x_0),\
\left.\left(\frac{\partial}{\partial x}\mp
f\right)p(x,t|x_0)\right| _{x=1/2\mp 1/2}=0,
\end{equation}
with the initial condition $\ p(x,0|x_0)=\delta(x-x_0)$. Equation
(\ref{fde1}) is solved by
\begin{equation}
p(x,z|x_0)=\left(e^{k|x-x_0|}+\kappa^2e^{2k-k|x-x_0|}+\kappa
e^{k(x+x_0)}+\kappa
e^{2k-k(x+x_0)}\right)/2k\left[\kappa^2e^{2k}-1\right],
\end{equation}
in the Laplace domain defined by
$g(z)=\int_0^{\infty}g(t)\exp(-zt)dt$. Here, $k\equiv\sqrt{z+f^2}$
and $\kappa=(k+f)/(k-f)$. To find the expression for the PDF
$P(x,y,t)$ we need the Laplace inverse of $p(x,z)$. Note that the
latter can be expanded in eigenmodes through
$p(x,t)=\sum_{n=0}^{\infty}
e^{\lambda_nt}\varphi_n(x)\varphi^*_n(x_0)$.

It is important to notice at this point that equation (\ref{fde1})
is {\em not} a physically meaningful 1D Fokker-Planck equation
(due to unphysical boundary conditions) and, indeed, it turns out
that for $f<0$ one of the $\lambda$'s, determined by the
transcendent equation $(\lambda+2f^2)\sinh\sqrt{\lambda+f^2}+2f
\sqrt{\lambda+f^2}\cosh\sqrt{\lambda+f^2}=0$, is actually
positive, i.e., gives a solution exponentially growing in time.
This type of behaviour is absent in the physical quantity
$P(x,y,t)$ due to the antisymmetrization procedure (\ref{sol}),
yet the $\lambda>0$ eigenmode plays an important role in the
Kramers type asymptotics obtained for $f\ll -1$. Details are
deferred to a future publication \cite{jonas_long}.

\section{Results for the coalescence time and position}

The coalescence time PDF, $\pi(t|x_0,y_0)$, can be found using the
conservation of probability
\begin{equation}
\int_0^t dt' \pi(t')+\int_0^1dy\int_0^y dx P(x,y,t)=1 ,
\end{equation}
where the two terms denote the probabilities that walkers have met
before time $t$ or have not, respectively. Then it follows from
the Fokker-Planck equation (\ref{fpe}) after integration by parts,
\begin{equation}\label{pdf}
\pi(t)= -\int_0^1dy\int_0^y dx \frac{\partial P(x,y,t)}{\partial
t}=\int_0^1dx\left.\left(\frac{\partial}{\partial y}-
\frac{\partial}{\partial
x}\right)P(x,y,t)\right|_{y=x}\equiv\int_0^1dx\varrho(x,t),
\end{equation}
where $\varrho(x,t)$ is the joint PDF for coalescence time and
meeting position of the two walkers. We are interested in $\pi(t)$
and its mean value (mean coalescence time)
$\tau=\int_0^{\infty}t\pi(t)dt$ together with the PDF of the
meeting position $\rho(x)=\int_0^{\infty}\varrho(x,t)dt$. By
standard identities of the Laplace transforms we note that
$\rho(x)= \varrho(x,z=0^+)$ and
$\tau=\left.-d\pi(z)/dz\right|_{z=0^+}$. The related analysis
becomes non-trivial as the \emph{auxiliary} Green's function
$p(x,t|x_0)$, as mentioned above, has a positive eigenvalue in the
case $f<0$. Details of the calculations using the theorem of
residues and high precision numerics will be presented in
reference \cite{jonas_long}.

\begin{figure}
\twoimages[width=4.77cm,angle=270]{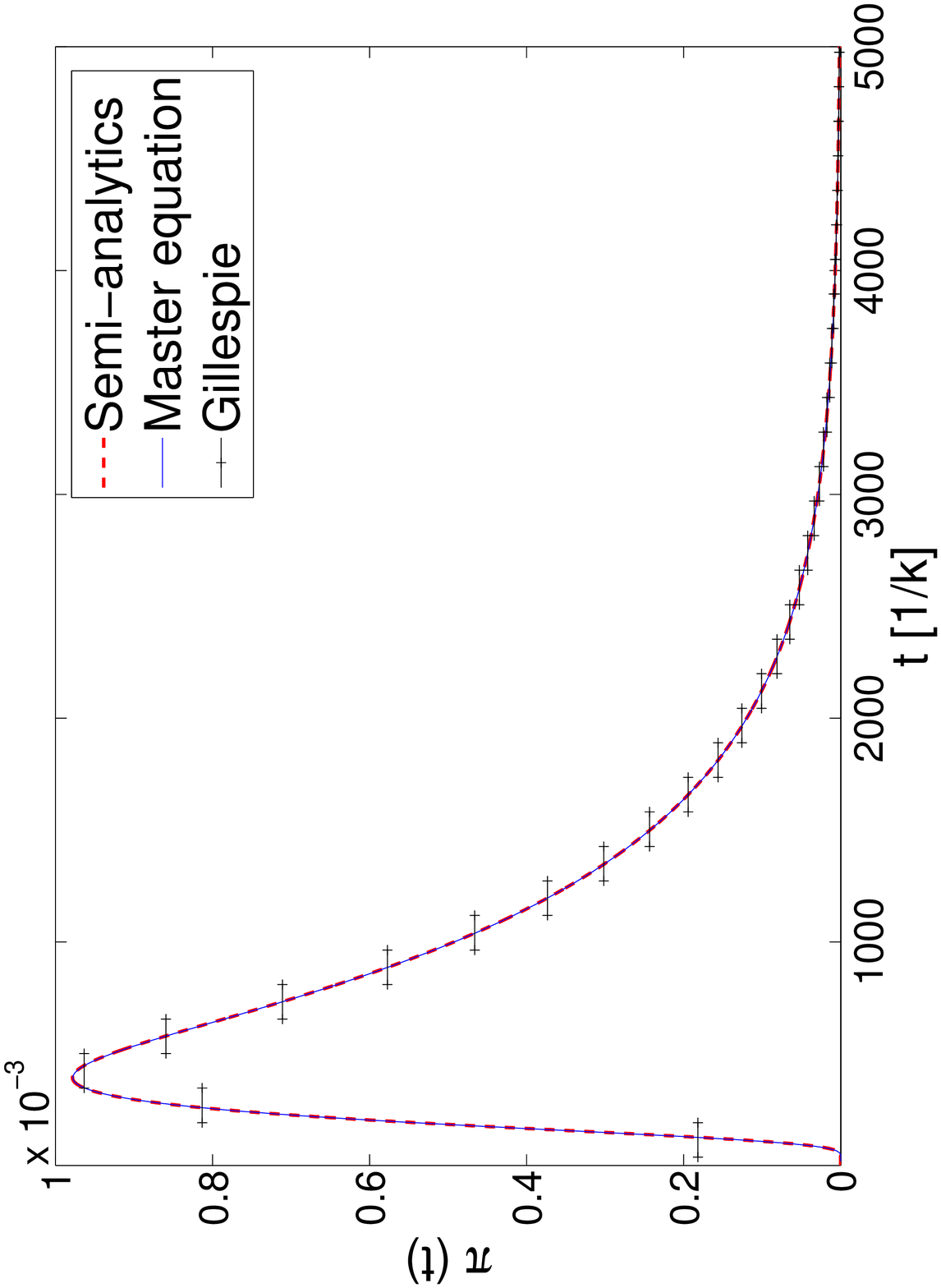}{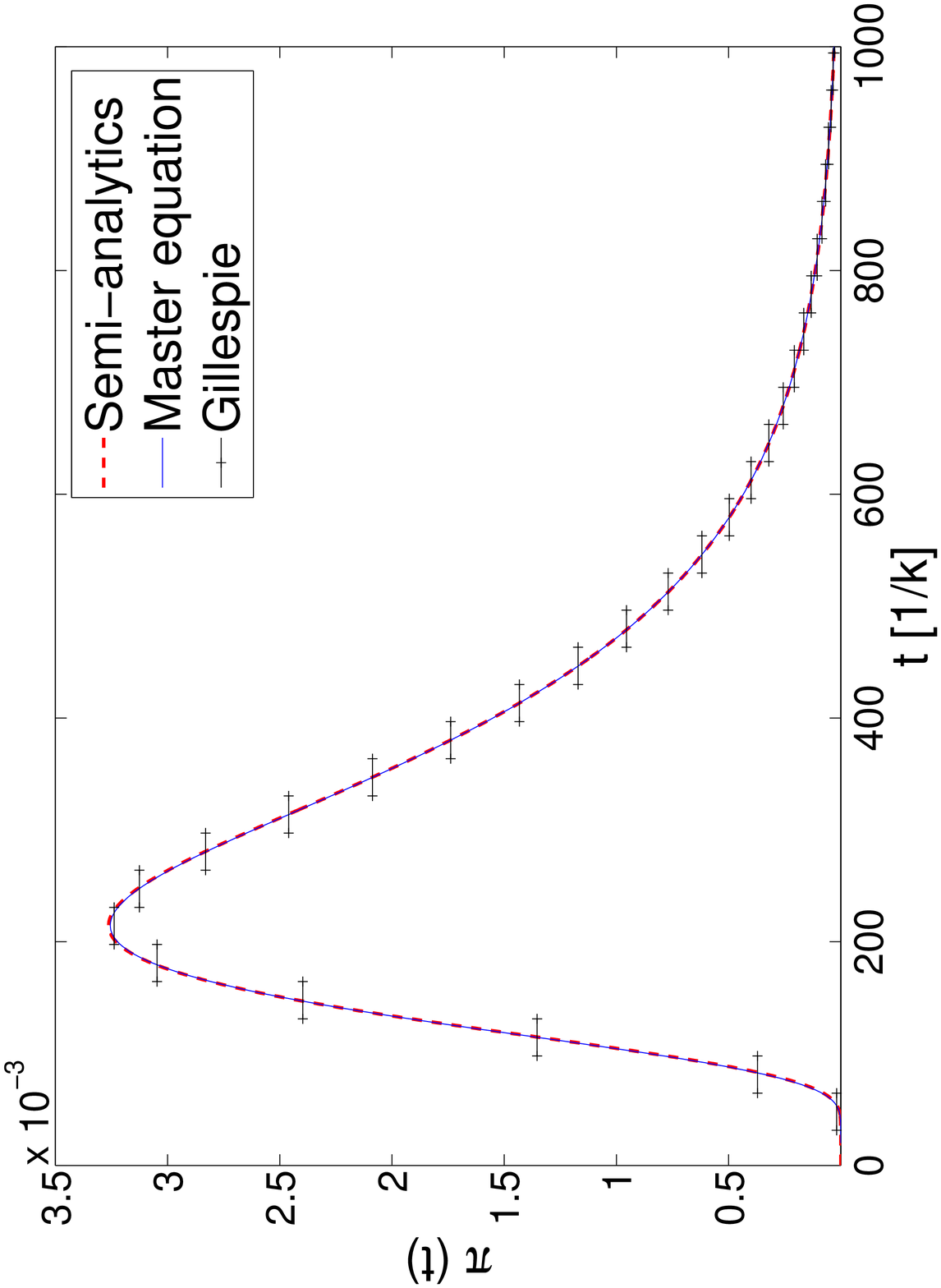}
\caption{Comparison of the results for the bubble coalescence time PDF
$\pi(t)$ obtained from the presented semi-analytical theory
(eqs.~(\ref{fpe})--(\ref{pdf})) by numerical inverse Laplace
transform, the master equation, and the stochastic simulation
(Gillespie), demonstrating excellent agreement. The values of the
parameters are the same as in figure \ref{trajs}, i.e., $u = 0.98$
(left) and $u=1.10$ (right), and $N=50$.} \label{comps}
\end{figure}

\begin{figure}
\twoimages[width=5cm,angle=270]{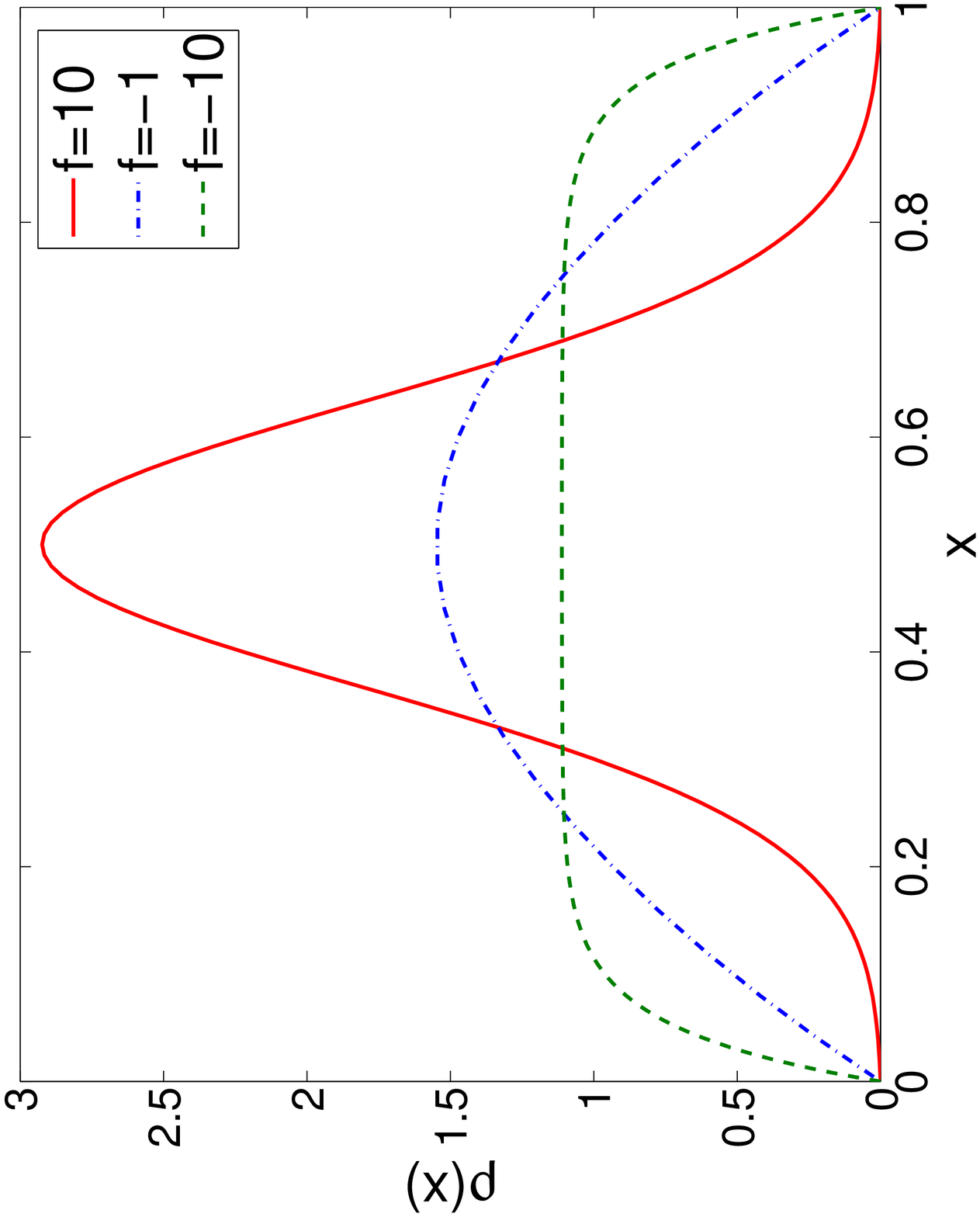}{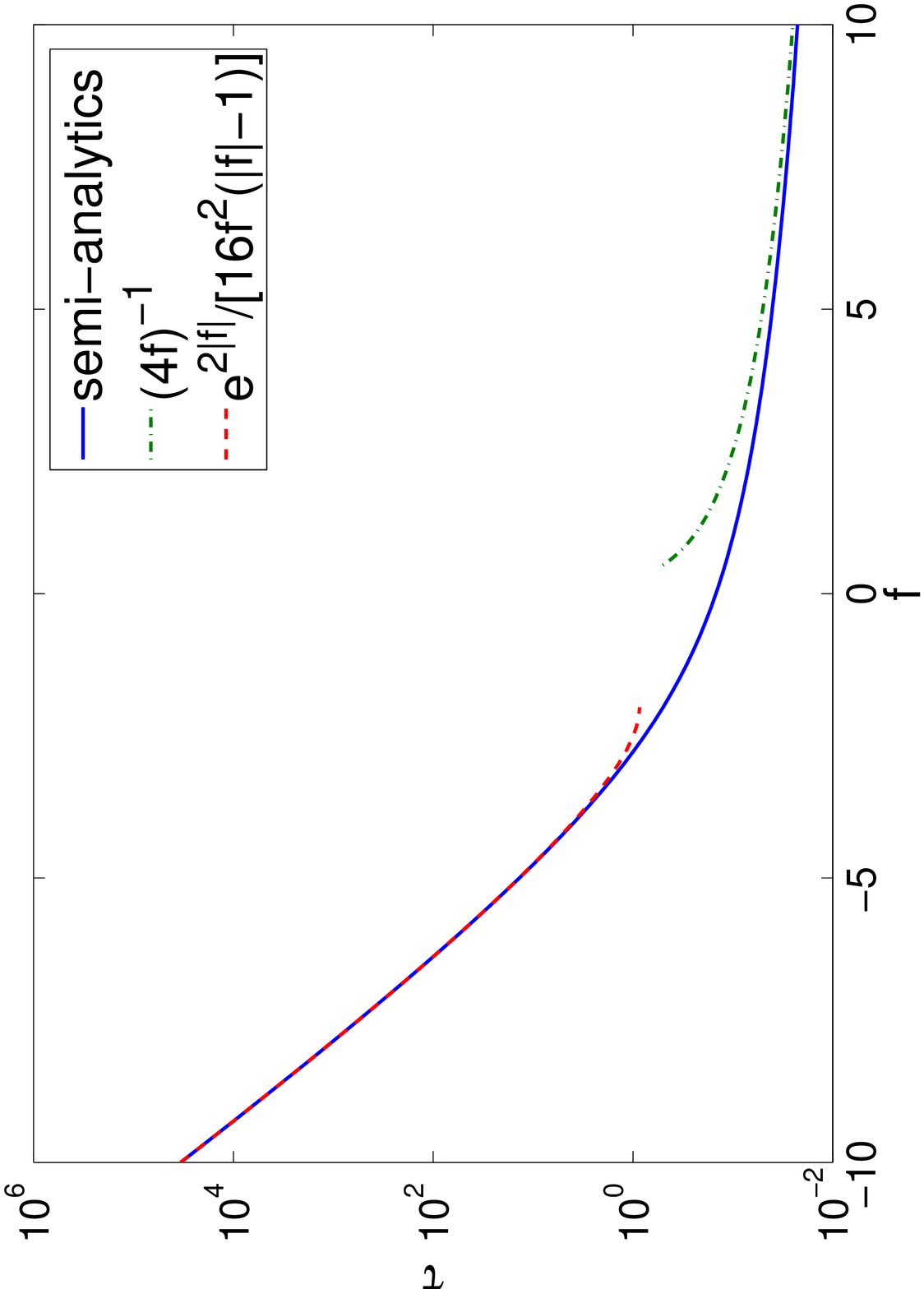} \caption{{\bf
Left:} PDF $\rho(x)$ of the coalescence position $x$ for various
values of the dimensionless force $f$. {\bf Right:} Mean
coalescence time $\tau$ (in dimensionless, i.e., scaled, units) as
a function of $f$. The two analytic asymptotic behaviours for
large and small $f$ are also shown.} \label{mfpt}
\end{figure}

Here, we discuss the final results for the coalescence time PDF
$\pi(t)$ in figure \ref{comps}, and for the mean coalescence time
$\tau$ and position PDF $\rho(x)$ in figure \ref{mfpt}. In all
cases the initial condition used is $x_0=0,\,y_0=1$, i.e., the
walkers start out at the reflecting walls or, in terms of the
original model, the bubbles start from the boundaries of the soft
zones. The coalescence time probability densities graphed in
figure \ref{comps} for two different values of $f$ (or $u$)
compare the results of all three mentioned methods (Gillespie,
master equation, and the semi-analytics) and show a sharp initial
raise of rather peaky structures followed by an exponential decay
for long times. The short-time increase stems from the initial
separation of the two forks that first have to go through several
random steps before having the chance to coalesce while the long
time asymptotics is fixed by the boundary conditions. The
agreement between the various computational approaches is very
good. The locations of the maximum probability peaks correlate
well with the termination points of sample trajectories depicted
in figure \ref{trajs} for the same parameters.
%For long times, the asymptotic behaviour is exponential, in both
%cases: While for the case $f<0$ corresponding to the crossing of a
%barrier ($T<T_b$) the exponential decay is characteristic of the
%Kramers-type behaviour, for the case $f\gg 1$ forcing the walkers
%towards the potential well in the centre ($T \gg T_b$) the
%exponential long-time asymptotics stems from the known first
%passage PDF of biased diffusion. Note, however, that in the former
%case the characteristic time scale is determined by the
%exponential wing of the PDF $\pi(t)$, in the latter case it
%corresponds to the maximum value of the curve.

The results for (i) $\rho(x)$ and (ii) $\tau$ calculated by the
semi-analytical method are shown in figure \ref{mfpt}. (i) The
curves for the PDF $\rho(x)$ of the coalescence position exhibit
a pronounced crossover from a relatively sharply peaked form to an
almost flat behaviour. The former occurs for large positive force
$f$, corresponding to a strong drift toward a potential well, with
negligible influence of the boundary conditions. In contrast, for
large negative $f$, corresponding to a high barrier for
coalescence, the insensitivity of $\rho(x)$ to the position $x$
can be explained in terms of a simple Arrhenius argument: The
probability of the walker to be at a position $x$ is proportional
to the Boltzmann weight, $\exp(-\beta\phi(x))$, where
$\phi(x)=-\int^xF(x')dx'$ is the free energy corresponding to the
force $F(x)$ (see figure \ref{sketch}). Then, the joint
probability to have both walkers meet at the same position is
given by the product $\exp(-\beta[\phi_L(x)+\phi_R(x)])\approx
\mathrm{const}$ as the two walkers are in opposite linear
potentials and the position dependence of the exponent cancels
out. This simple picture necessarily breaks down close to the
boundaries. (ii) The $f$-dependence of the mean first passage time
$\tau$ crosses over from the $\tau\simeq1/f$ behaviour typical for
diffusion in a strong positive force pushing the two walkers
together, to the exponential form $\tau\simeq\exp(2|f|)$ of the
associated Kramers problem. The former problem was studied in
reference \cite{hame} by neglecting the boundaries and switching
to the relative coordinate description which enables one to find
the analytic result $\tau=1/(4f)$. For the Kramers problem ($f\ll
-1$) the analytic solution for both
$\rho(x)=[1-e^{-2|f|x}-e^{-2|f|(1-x)}]|f|/(|f|-1)$ and
$\tau=e^{2|f|}/[16f^2(|f|-1)]$ can be found rather easily
\cite{jonas_long} by the expansion into the lowest two eigenmodes
of $p(x,t|x_0)$.

\section{Conclusions}

%While the double-helical structure of DNA shields the nucleobases from
%damage and therefore preserves the genetic information, the infrequent
%events of Watson-Crick bp-opening play a crucial role in the function
%of DNA, in particular, for the binding of proteins and chemicals
%\cite{kornberg}. Quantitative knowledge of the nature of DNA breathing
%is essential for a better understanding of the molecular biology
%of DNA-based interactions like gene regulation or signalling.
%This offers the possibility to study in detail the fundamental
%kinetics underlying DNA breathing but also to measure at high
%precision the base stacking free energies.It will be interesting
%to analyze in more detail the dependence on heteropolymeric
%barrier sequences, or in the presence of proteins that selectively
%bind to single-stranded DNA and therefore enhance the dissociation
%of bps.
%The exact form of the Laplace transform of the latter was derived,
%while numerical analysis revealed the behaviour in the time domain.

In the present study we investigate two-bubble coalescence in a
designed DNA construct consisting of two soft regions separated by
a more stable barrier zone.  We present a continuous
semi-analytical theory yielding the coalescence time and position
PDF's and show that the results agree well with the numerical results
obtained with the discrete master equation and Gillespie
stochastic simulation schemes. We note that for long barriers,
whose sequence is arranged as random energy landscape, the
propagation of the zipping forks may become subdiffusive in
time\cite{hwa} and the associated first passage time process is
generated from the present results with the help of the
subordination formalism \cite{report1}. The mathematical analysis
reduces to the previously unaddressed case of two vicious walkers
in \emph{opposite\/} linear potentials. We demonstrate that its
solution can be constructed by antisymmetrization of appropriate
auxiliary one-variable densities.

\acknowledgments

The work of T.~N.\ is a part of the research plan MSM 0021620834
financed by the Ministry of Education of the Czech Republic.
R.~M.\ acknowledges the Natural Sciences and Engineering Research
Council (NSERC) of Canada, and the Canada Research Chairs
programme, for support. This work was started at CPiP 2005
(Computational Problems in Physics, Helsinki, May 2005) supported
by NordForsk, Nordita, and Finnish NGSMP.

\end{document}

%% file: bubbsketch_new_thin.pstex_t
\begin{picture}(0,0)%
\includegraphics{bubbsketch_new_thin.pstex}%
\end{picture}%
\setlength{\unitlength}{4144sp}%
\begingroup\makeatletter\ifx\SetFigFont\undefined%
\gdef\SetFigFont#1#2#3#4#5{%
  \reset@font\fontsize{#1}{#2pt}%
  \fontfamily{#3}\fontseries{#4}\fontshape{#5}%
  \selectfont}%
\fi\endgroup%
\begin{picture}(8965,3828)(2461,-4159)
\put(8011,-1951){\makebox(0,0)[lb]{\smash{{\SetFigFont{12}{14.4}{\rmdefault}{\mddefault}{\updefault}{\color[rgb]{0,0,0}\fbox{$T_s <T$}}%
}}}}
\put(6661,-466){\makebox(0,0)[lb]{\smash{{\SetFigFont{12}{14.4}{\rmdefault}{\mddefault}{\updefault}{\color[rgb]{0,0,0}$N_R$}%
}}}}
\put(5221,-466){\makebox(0,0)[lb]{\smash{{\SetFigFont{12}{14.4}{\rmdefault}{\mddefault}{\updefault}{\color[rgb]{0,0,0}$N$}%
}}}}
\put(2476,-1996){\makebox(0,0)[lb]{\smash{{\SetFigFont{12}{14.4}{\rmdefault}{\mddefault}{\updefault}{\color[rgb]{0,0,0}$b_1$)}%
}}}}
\put(2476,-2941){\makebox(0,0)[lb]{\smash{{\SetFigFont{12}{14.4}{\rmdefault}{\mddefault}{\updefault}{\color[rgb]{0,0,0}$b_2$)}%
}}}}
\put(2476,-3931){\makebox(0,0)[lb]{\smash{{\SetFigFont{12}{14.4}{\rmdefault}{\mddefault}{\updefault}{\color[rgb]{0,0,0}$b_3$)}%
}}}}
\put(8056,-3841){\makebox(0,0)[lb]{\smash{{\SetFigFont{12}{14.4}{\rmdefault}{\mddefault}{\updefault}{\color[rgb]{0,0,0}\Large time}%
}}}}
\put(3691,-466){\makebox(0,0)[lb]{\smash{{\SetFigFont{12}{14.4}{\rmdefault}{\mddefault}{\updefault}{\color[rgb]{0,0,0} $N_L$}%
}}}}
\put(7966,-1006){\makebox(0,0)[lb]{\smash{{\SetFigFont{12}{14.4}{\rmdefault}{\mddefault}{\updefault}{\color[rgb]{0,0,0}\fbox{$T<T_s <T_b$}}%
}}}}
\put(10936,-3621){\makebox(0,0)[lb]{\smash{{\SetFigFont{12}{14.4}{\rmdefault}{\mddefault}{\updefault}{\color[rgb]{0,0,0}$x,y$}%
}}}}
\put(4816,-3481){\makebox(0,0)[lb]{\smash{{\SetFigFont{12}{14.4}{\rmdefault}{\mddefault}{\updefault}{\color[rgb]{0,0,0}$X$}%
}}}}
\put(5771,-3481){\makebox(0,0)[lb]{\smash{{\SetFigFont{12}{14.4}{\rmdefault}{\mddefault}{\updefault}{\color[rgb]{0,0,0}$Y$}%
}}}}
\put(2566,-1006){\makebox(0,0)[lb]{\smash{{\SetFigFont{12}{14.4}{\rmdefault}{\mddefault}{\updefault}{\color[rgb]{0,0,0}$a$)}%
}}}}
\put(10026,-3056){\makebox(0,0)[lb]{\smash{{\SetFigFont{12}{14.4}{\rmdefault}{\mddefault}{\updefault}{\color[rgb]{0,0,0}$-2fx$}%
}}}}
\put(10131,-1396){\makebox(0,0)[lb]{\smash{{\SetFigFont{12}{14.4}{\rmdefault}{\mddefault}{\updefault}{\color[rgb]{0,0,0}$2f(1-y)$}%
}}}}
\put(4421,-3481){\makebox(0,0)[lb]{\smash{{\SetFigFont{12}{14.4}{\rmdefault}{\mddefault}{\updefault}{\color[rgb]{0,0,0}$0$}%
}}}}
\put(6071,-3481){\makebox(0,0)[lb]{\smash{{\SetFigFont{12}{14.4}{\rmdefault}{\mddefault}{\updefault}{\color[rgb]{0,0,0}$N$}%
}}}}
\end{picture}%